\documentclass{acm_proc_article-sp-tweeked}

\usepackage[hyphens]{url}
\usepackage[boxed]{algorithm}
\usepackage{clrscode3e}
\usepackage{afterpage}
\usepackage{listings}
\usepackage{multirow}

\newcommand{\bigImageSize}{0.8\textwidth}

\begin{document}

%\title{A Sample {\ttlit ACM} SIG Proceedings Paper in LaTeX
%Format\titlenote{(Does NOT produce the permission block, copyright information nor page numbering). For use with ACM\_PROC\_ARTICLE-SP.CLS. Supported by ACM.}}
\title{Avoiding Spoilers in Fan Wikis of Episodic Fiction}
%\subtitle{[Extended Abstract]
%\titlenote{A full version of this paper is available as
%\textit{Author's Guide to Preparing ACM SIG Proceedings Using
%\LaTeX$2_\epsilon$\ and BibTeX} at

\numberofauthors{2} %  in this sample file, there are a *total*
\author{
% 1st. author
\alignauthor
Shawn M. Jones\\
       \affaddr{Old Dominion University}\\
       \affaddr{Norfolk, VA, USA}\\
       \email{sjone@cs.odu.edu}
% 2nd. author
\alignauthor
Michael L. Nelson\\
       \affaddr{Old Dominion University}\\
       \affaddr{Norfolk, VA, USA}\\
       \email{mln@cs.odu.edu}
%% 3rd. author
%\alignauthor Lars Th{\o}rv{\"a}ld\titlenote{This author is the
%one who did all the really hard work.}\\
%       \affaddr{The Th{\o}rv{\"a}ld Group}\\
%       \affaddr{1 Th{\o}rv{\"a}ld Circle}\\
%       \affaddr{Hekla, Iceland}\\
%       \email{larst@affiliation.org}
%\and  % use '\and' if you need 'another row' of author names
%% 4th. author
%\alignauthor Lawrence P. Leipuner\\
%       \affaddr{Brookhaven Laboratories}\\
%       \affaddr{Brookhaven National Lab}\\
%       \affaddr{P.O. Box 5000}\\
%       \email{lleipuner@researchlabs.org}
%% 5th. author
%\alignauthor Sean Fogarty\\
%       \affaddr{NASA Ames Research Center}\\
%       \affaddr{Moffett Field}\\
%       \affaddr{California 94035}\\
%       \email{fogartys@amesres.org}
%% 6th. author
%\alignauthor Charles Palmer\\
%       \affaddr{Palmer Research Laboratories}\\
%       \affaddr{8600 Datapoint Drive}\\
%       \affaddr{San Antonio, Texas 78229}\\
%       \email{cpalmer@prl.com}
}
% There's nothing stopping you putting the seventh, eighth, etc.
% author on the opening page (as the 'third row') but we ask,
% for aesthetic reasons that you place these 'additional authors'
% in the \additional authors block, viz.
%\additionalauthors{Additional authors: John Smith (The Th{\o}rv{\"a}ld Group,
%email: {\texttt{jsmith@affiliation.org}}) and Julius P.~Kumquat
%(The Kumquat Consortium, email: {\texttt{jpkumquat@consortium.net}}).}
% Just remember to make sure that the TOTAL number of authors
% is the number that will appear on the first page PLUS the
% number that will appear in the \additionalauthors section.

\maketitle
\begin{abstract}

A variety of fan-based wikis about episodic fiction (e.g., television shows, novels, movies) exist on the World Wide Web.  These wikis provide a wealth of information about complex stories, but if readers are behind in their viewing they run the risk of encountering ``spoilers'' -- information that gives away key plot points before the intended time of the show's writers.  Enterprising readers might browse the wiki in a web archive so as to view the page prior to a specific episode date and thereby avoid spoilers. Unfortunately, due to how web archives choose the ``best'' page, it is still possible to see spoilers (especially in sparse archives).

In this paper we discuss how to use Memento to avoid spoilers.  Memento uses TimeGates to determine which best archived page to give back to the user, currently using a minimum distance heuristic.  We quantify how this heuristic is inadequate for avoiding spoilers, analyzing data collected from fan wikis and the Internet Archive.  We create an algorithm for calculating the probability of encountering a spoiler in a given wiki article.  We conduct an experiment with 16 wiki sites for popular television shows.  We find that 38\% of those pages are unavailable in the Internet Archive.  We find that when accessing fan wiki pages in the Internet Archive there is as much as a 66\% chance of encountering a spoiler.  Using sample access logs from the Internet Archive, we find that 19\% of actual requests to the Wayback Machine for \url{wikia.com} pages ended in spoilers.  We suggest the use of a different minimum distance heuristic, minpast, for wikis, using the desired datetime as an upper bound.

%A variety of fan-based wikis exist on the World Wide Web for fans of episodic fiction.  Unfortunately, these sources have also led to a rise in spoilers as fans improve these resources with more current information.  The Memento Framework can be used to avoid spoilers.  The Memento Framework uses constructs called TimeGates to redirect users to past versions of web pages using a date that the user supplies.  Currently, most TimeGates utilize a minimum distance heuristic to decide to which past web page the user is redirected, which makes sense for web archives because they are sparse.  We document and suggest the use of a different heuristic, minimum distance without going over the date supplied by the user.  Our proposed heuristic is best used on wikis because they retain every revision of a given page.

\end{abstract}

%\category{H.3.5}{Information Storage and Retrieval}{Online Information Services}
%A category including the fourth, optional field follows...
%\category{D.2.8}{Software Engineering}{Metrics}[complexity measures, performance measures]

%\terms{Digital Preservation, HTTP, Resource Versioning, Web Archiving, Wikis, Spoilers}

%\keywords{ACM proceedings, \LaTeX, text tagging} % NOT required for Proceedings


\section{Introduction}

From \emph{How I Met Your Mother} to \emph{Game of Thrones}, fans have created fan-based wikis based on their favorite episodic fiction.  For a community of fans these wikis become the focal point for continued discussion and documenting the details of the fictional milieu.  The first study on fan wikis was done on the wiki \emph{Lostpedia}, for the show \emph{Lost} \cite{Mittell2009}.

Unfortunately, due to the rise in the availability of recorded media, viewing television shows after their air date has become more common, making the use of these wikis difficult for those who have not yet consumed all of the episodes to date, leading to \emph{spoilers}.  Spoilers are defined as pieces of information that user wants to control the time and place of their consumption, preferring to consume them in the order that the author (or director) intended. If these pieces of information are delivered in the wrong order, enjoyment about a movie or television program is damaged \cite{Jeon2013}.  The problem of spoilers has been reported in popular media for years, from such sources as \emph{CNN} \cite{Gross2014} and \emph{The New York Times}  \cite{Hart2009}.

The Memento Framework \cite{VandeSompel2009, memento-rfc:2013} can be used to avoid spoilers on the web \cite{blog-avoid-spoilers, JonesThesis2015}.  Memento allows one to extend content negotiation into the dimension of time, a process called \textbf{datetime negotiation}, allowing a user to choose a date prior to the episode they have not seen and view the web as it looked at that time.

%The typical practice of using \emph{spoiler alert} notices causes users to avoid these fan sites, potentially costing advertising revenue \cite{Tsang2009}. Using Memento to avoid spoilers can allowing their website visitors to safely visit from these sites, and receive advertisements.

%We quantify why the Memento MediaWiki Extension is needed, and why existing technologies, such as the \emph{Wayback Machine}, created by the Internet Archive, and many existing Memento TimeGates, can not be used to reliably avoid spoilers.  Currently, both the Wayback Machine and Memento TimeGates utilize a minimum distance heuristic, which we define here as \textbf{mindist}, to find the best memento to present to a user.  We show not only that mindist can lead to spoilers, but also the probability of encountering spoilers when using mindist with fan wikis for several popular television shows.

%\subsection{Wiki Revisions, Memento, and Spoilers}

Memento provides several resource types that play a role in datetime negotiation.  First, the original resource, also noted as a URI-R, is the page for which we want the past version. In MediaWiki parlance, it is called a topic URI, and refers to the wiki article in its current state.  Then we have the memento, from which the Memento Framework gets its name, also noted as URI-M. It is the past version of the page. In MediaWiki parlance, it is called a oldid page.  Third, we have the TimeMap, also noted as URI-T, which is a resource associated with the original resource from which a list of mementos for that resource are available. The TimeMap provides a list of URI-Ms and datetimes in a well-defined format, but does not contain any article content.  Finally, we have the TimeGate, also noted as URI-G, which is the resource associated with the original resource that provides datetime negotiation. It is the URI to which the user sends a datetime and receives information about which memento (URI-M) is the best match for that datetime. The TimeGate only processes and redirects; it provides no representations itself.

Wikis preserve every revision of a page as mementos, accessible via a series of URI-Ms.  The web archive then captures some of those revisions as general mementos, accessible via a different series of URI-Ms.  Unfortunately, for a web archive, there are \textbf{missed updates} that are never recorded, so we are unsure of the interval for which any given general memento is valid.  For a wiki, we have every revision and \emph{no missed updates}, so we do know the interval of their validity.  This makes wiki revisions a special case of mementos.  For the sake of this paper, to differentiate between the two sources, we use the term \textbf{revision} when referring to a memento saved by a wiki, and the term \textbf{memento} when referring to the more general mementos residing at a web archive.  The full discussion of models and durations of validity is outside of the scope of this paper. 

%Figure \ref{fig:memento-wiki-caps} shows two timelines.  The bottom timeline consists of mementos captured by a web archive.  The top line consists of several revisions of a wiki page.  From this figure, we see that memento $m_k$ was archived by the web archive at datetime $t_{14}$, which we denote as $m_k@t_{14}$.  Likewise, revision $r_{j-4}@t_2$ denotes that wiki revision $r_{j-4}$ was saved at time $t_2$.  Arrows between the memento line and the revision line show which mementos are captures of which revisions.  We denote this as $m_k \equiv r_j$, indicating that $m_k$ is a capture of $r_j$.  We see that revisions $r_{j-3}$, $r_{j-2}$, and $r_{j-1}$ are never captured, making them missed updates.

\begin{figure}[h!t]
\centering
	\includegraphics[width=0.45\textwidth]{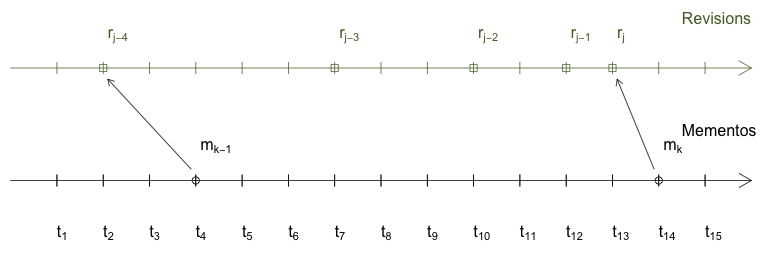}
	\caption{Example Timeline Showing Captured Mementos of Wiki Edits}	
	\label{fig:memento-wiki-caps}

    \includegraphics[width=0.45\textwidth]{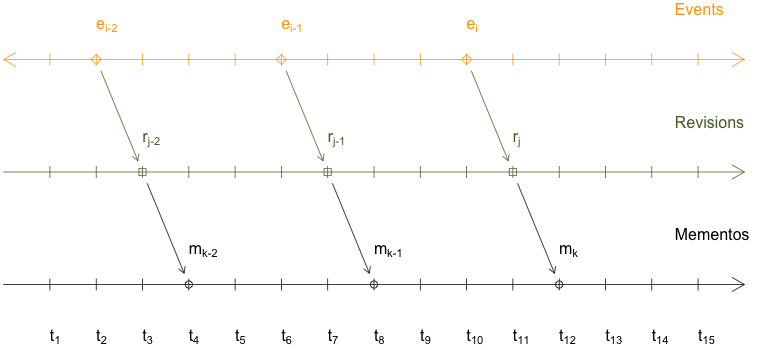}
    \caption{Each event can inspire a new wiki revision which may be captured as a memento by a web archive}    
    \label{fig:event-representation-memento}

    \includegraphics[width=0.45\textwidth]{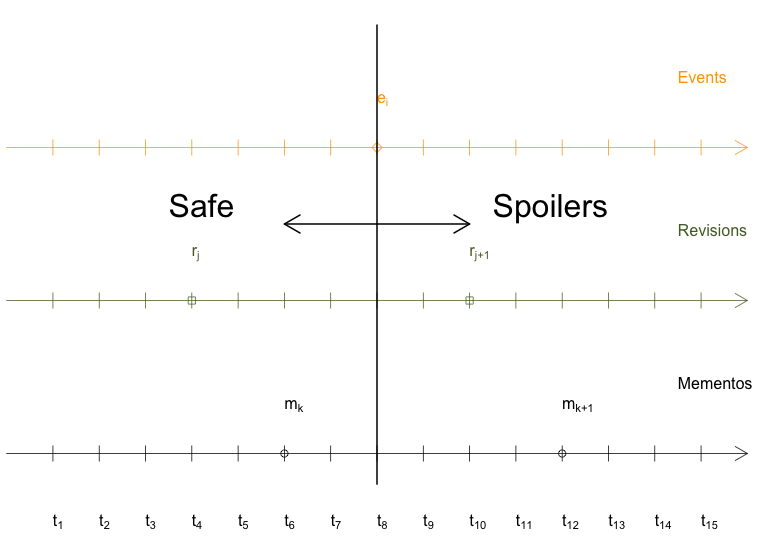}
    \caption{Representation of our Na\"{i}ve Spoiler Concept}    
    \label{fig:naive-spoiler}
\end{figure}

Figure \ref{fig:memento-wiki-caps} shows two timelines.  The bottom timeline consists of mementos captured by a web archive.  The top line consists of several revisions of a wiki page.  From this figure, we see that memento $m_k$ was archived by the web archive at datetime $t_{14}$, which we denote as $m_k@t_{14}$, or, also more generically $t_{m_k}$.  Likewise, revision $t_{r_{j-4}}@t_2$ denotes the time for which wiki revision $r_{j-4}$ was saved.  Arrows between the memento line and the revision line show which mementos are captures of which revisions.  We denote this as $m_k \equiv r_j$, indicating that $m_k$ is a capture of $r_j$.  We see that revisions $r_{j-3}$, $r_{j-2}$, and $r_{j-1}$ are never captured, making them missed updates.

In Figure \ref{fig:event-representation-memento} we add a third timeline for events, showing the pattern observed by Steiner \cite{Steiner:2013:MNM:2487788.2488049} where events inspire wiki revisions to be created.   In this case events correspond to television episodes.  As seen above, these edits are eventually captured by the web archive.  We use the nomenclature $t_{e_i}$ to refer to the time of the $i^{th}$ episode.  We also use the $e_1$ to refer to the first episode and $e_n$ as the latest (or last) episode.

For this paper, we define the term \textbf{spoiler} na\"{i}vely as any memento that exists after the date desired by the user, regardless of the content of the memento.  Figure \ref{fig:naive-spoiler} illustrates this concept using $e_i$ as the episode datetime, and $r_j$ and $r_{j+1}$ as revisions on either side of this event.  Based on our definition revision $r_j$ is \textbf{safe} because it exists prior to episode $e_i$ that the user is trying to avoid.  It is assumed that revision $r_{j+1}$ contains spoilers because that wiki edit occurred after the $e_i$.

It is this relationship between events and revisions that allow our spoiler solution to work.  Fans who edit wiki pages typically have no knowledge of an episode's content until that episode airs, meaning that revisions containing that information must come after the episode.

%\subsection{Structure of this Paper}

In determining the best memento to which the user should be directed, web archives use a minimum distance heuristic.  We demonstrate that this heuristic is not useful for avoiding spoilers. Fan wikis are a special case, because they are updated frequently and many of their users want to avoid spoilers. We do not seek to change the Internet Archive's processes. We can use Memento directly on wikis to avoid spoilers because wikis have access to all revisions \cite{Leuf2001}.  These revisions are mementos in their own right, and because we have all revisions, we can use a different heuristic (minpast), that avoids mementos after the date requested by the user.  Thus, by using Memento directly on a wiki, one can avoid spoilers in fan wikis.  The Memento MediaWiki Extension provides this functionality for MediaWiki \cite{Jones2014}, allowing spoiler avoidance for those who install the extension.  
 
As part of this temporal analysis, we will further define the two heuristics under consideration, mindist and minpast.  We will also show that there is a 66\% probability of encountering a spoiler when using web archives to access prior versions of wikis, because web archives use mindist.  In addition to not reliably helping users avoid spoilers, we find that 38\% of the pages in our sample are not available in the archive.

In this paper, we briefly show what others have done to study the spoiler problem, then we discuss what previous studies have been done on wikis.  From here we discuss two different TimeGate heuristics and how minpast is preferred over mindist when a user is trying to avoid spoilers.  Then we discuss how the mindist heuristic can lead to spoiler areas, where a user selects a datetime prior to an episode they want to avoid, but are still directed into the future.  Using these spoiler areas, we then show how one can calculate the probability of encountering a spoiler for a given page.

Armed with these concepts, we show the results of a study performed on 16 fan wikis for popular television shows \cite{JonesThesis2015}, showing not only that these spoiler areas exist for users, but also the probabilities of encountering spoilers in these sites.

We then discuss the results of a second study on logs from the Wayback Machine, showing that 19\% of all requests end up in the future, indicating that the spoiler problem is real, and that the Wayback Machine is not a reliable tool for avoiding spoilers on the web.

\section{Related Work}

Schirra, Sun, and Bently conducted a study of two-screen viewing while the television show \emph{Downton Abbey} was airing \cite{Schirra2014}. Two-screen viewing is a process whereby those watching a television show episode discuss the show on a social media web site, such as Twitter, while the episode is airing.  A similar study was conducted by Johns \cite{Johns2012}.  Both studies discovered that users would use elaborate methods to avoid revealing and encountering spoilers in social media as well as the current versions of web sites.

%This study consisted of a sample of 2,234 participants who live-tweeted during the highly anticipated third season premier and beyond.  Some of the United Kingdom live-tweeters would hold off revealing spoilers, but still live-tweet with their American friends so that they could vicariously share in the story reveals and plot information as the Americans experienced it.  Others would concoct methods to communicate major plot twists, such as using ambiguous pronouns, without spoiling the story for their friends.  This is also consistent with a study conducted by Johns, also using interviews in a small group of participants who also engaged in \emph{two screen viewing}, a more generic name for live-tweeting \cite{Johns2012}.  Both studies discovered that those who used digital video recording (DVR) devices, such as the TiVo, would avoid social media until they had watched their show.  Also, some would eschew DVRs because they wanted to participate in live-tweeting.

Because of the phenomenon of spoilers in social media, Boyd-Graber, Glasgow, and Zajac conducted an evaluation of machine learning approaches to find spoilers in social media posts \cite{Boyd-graber2013}.  They used classifiers on multiple sources to determine which posts should be blocked.  They mention that spoilers refer to events ``\emph{later} than the viewer's knowledge of the current work'', suggesting that any machine learning technique used for avoiding spoilers in social media must be smarter than just blocking all posts about a particular topic \cite{Golbeck2012, Jeon2013}. Inspired by this work were software packages that block spoilers from a user's social media feed, such as Spoiler Shield \cite{pickingCarrots} and the Netflix Spoiler Foiler \cite{Denham2014}.

%Their classifiers were trained by crowdsourcing and mining data from the Internet Movie Database\footnote{http://imdb.com}, TV Tropes\footnote{http://tvtropes.org}, and Episode Guides\footnote{http://epguides.com} online resources.  By utilizing these additional sources, they were able to use machine learning techniques to identify spoilers better than their predecessors, who relied primarily on term matching and small data sets. 

We are proposing an orthogonal concept relating to fan wikis, not social media.  We are also not blocking resources, rather indicating that the fan wiki pages can still be useful resources if past versions of them are accessible to users.  Our solution can be combined with a content-based approach, but we are proposing a structural solution that can be combined with content-based solutions in the future.

Almedia, Mozafari, and Cho produced one of the first studies of the behavior of contributors to Wikipedia \cite{Almeida2007}.  The authors discover that there are distinct groups of Wikipedia contributors.  They suggest that as the number of articles increases, the contributors' attention is split among more and more content, resulting in the larger number of revising contributors rather than article creators.  This informs our notion of number of edits as a surrogate to the popularity of a page.

%Vong, Lim, Sun, et al. have developed models of evaluating Wikipedia articles so they can be flagged as \emph{controversial} \cite{Vuong:2008:RCW:1341531.1341556}. This way editors can focus their efforts on resolving controversies in particular articles, but also allowing others to see which controversial topics in Wikipedia are indicative of the real world controversies, allowing for further areas of study.

%In 2010, Lucassen and Schraggen again evaluated the ``trustworthiness'' of Wikipedia articles \cite{Lucassen:2010:TWU:1772938.1772944}.  They discuss how, by 2010, Wikipedia contains an Editorial Team that evaluates article quality and flags those articles that are considered to be of good quality and those that need work.  Their contribution is a series of features that indicate how Wikipedia users evaluate articles.  These features can then be used in the future for further evaluation by experts.

%We highlight these studies to indicate that there has been a lot of study on what Wikipedia can be.  The fan-based wikis we are attempting to avoid spoilers in tend to be central hubs of activity for those seeking to find information on their favorite fiction.  Wikipedia has undergone an evolution from completely closed to completely open to now having recommendations of articles by committee.  The wiki fan sites that we have reviewed are in various stages of this evolution, depending on how large a user base they have.

Additionally, there has been some effort of preserving wiki pages outside of the Internet Archive.  Popitsch, Mosser, and Phillipp have created the UROBE project for archiving wiki representations in a generic format that can then be reconstituted into many other formats for data analysis \cite{Popitsch2010}.  Interestingly, they anticipate attaching their process to Memento at some point later in their research so that past versions of their archives can be accessed by datetime.  

%As of the paper's publication, they were only preserving the content of wikis externally, albeit via a different method.

\section{Memento TimeGate Heuristics}

When the user selects a desired datetime prior to the episode they have not yet seen, the TimeGate is what determines which memento they are redirected to.  In the case of spoilers, the wrong heuristic can redirect the user to a spoiler even though they requested a datetime prior to the event that would have caused the spoiler.  

%Because of this possibility, we identify the mindist and minpast heuristics for use with Memento TimeGates and why minpast is preferred when avoiding spoilers in wikis, but also why mindist is preferred when browsing sparse web archives.

\begin{figure}[t]

%\caption{Demonstration of the \emph{mindist} heuristic, in this case $m_2@t_7$ is chosen because it is closest to $t_a$}
%\centering
%\includegraphics[width=0.5 \textwidth]{images/mindist-example-1.png}
%\label{fig:tg-mindist-diagram-1}

%\caption{Demonstration of the \emph{mindist} heuristic; in this case $m_3@t_{10}$ is chosen because it is closest to $t_a$}
%\centering
%\includegraphics[width=0.5\textwidth]{images/mindist-example-2.png}
%\label{fig:tg-mindist-diagram-2}

\centering
\includegraphics[width=0.45\textwidth]{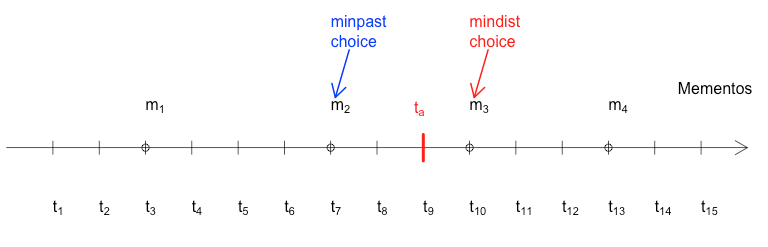}
\caption{Demonstration of the \emph{mindist} and \emph{minpast} heuristics; $m_3@t_{10}$ is chosen by mindist whereas $m_2@t_7$ is chosen by minpast}
\label{fig:tg-mindist-minpast}

\end{figure}

Memento TimeGates accept two arguments from the user: desired datetime (specified in the Accept-Datetime header) and a URI-R; and they return the \emph{best} URI-M using some heuristic \cite{Ainsworth2014}.  RFC 7089 leaves the heuristic of finding the \emph{best} URI-M up to the implementor, stating that ``the exact nature of the selection algorithm is at the server's discretion but is intended to be consistent'' \cite{memento-rfc:2013}.  Figure \ref{fig:tg-mindist-minpast} shows the differences between the mindist and minpast heuristics used for TimeGates.

%\subsection{Mindist}

\textbf{Mindist} (minimum distance) finds the closest memento to the given desired datetime $t_a$.  Mindist is best used for web archives, which are typically sparse, meaning they may have missed many revisions of a page.  In this case, a user would want the closest memento they can get to the date they are requesting because the dates of capture may be wildly distant from one another.  Because of the fact that it may choose mementos from a date after the desired datetime, mindist is not a reliable heuristic for avoiding spoilers.

This heuristic is useful in cases where there are few mementos recorded for a web page.  Consider an example where only two mementos exist, from 2003 and 2009.  If the user wishes to see the page as it looked on 2008, the 2009 (minimum distance) memento is likely best.  Most web archives are sparse, hence mindist is used to satisfy the majority of use cases.  This heuristic is what the Wayback Machine uses, and is not user-configurable.

\textbf{Minpast}, short for minimum distance in the past, finds the closest memento to the desired datetime $t_a$, but without going over $t_a$.  Minpast is best used for archives are abundant with mementos.  Ideally, minpast should be used if every revision of a resource has been archived, as with wikis.  For wikis, the value of desired datetime $t_a$ corresponds to a revision that actually existed at the time of $t_a$.  For web archives that are not abundant, information may be lost because they may not have captured all revisions.  Minpast can be used to avoid spoilers.  If we select a value for $t_a$ prior to the event we want to avoid, then minpast will not find any mementos after $t_a$.  It is best used for wikis where we have access to all revisions because we can definitively state that the memento returned is the page as it existed at $t_a$.

\section{Spoiler Probabilities}

By studying mindist using wiki revisions and the mementos corresponding to them, we can measure the probability of encountering a spoiler for a given wiki page in web archives.

The set of datetimes where the user is redirected to a memento after the episode, even though they chose a datetime prior to the episode is defined as a \textbf{spoiler area}.

\begin{figure}[t]
	\includegraphics[width=0.48\textwidth]{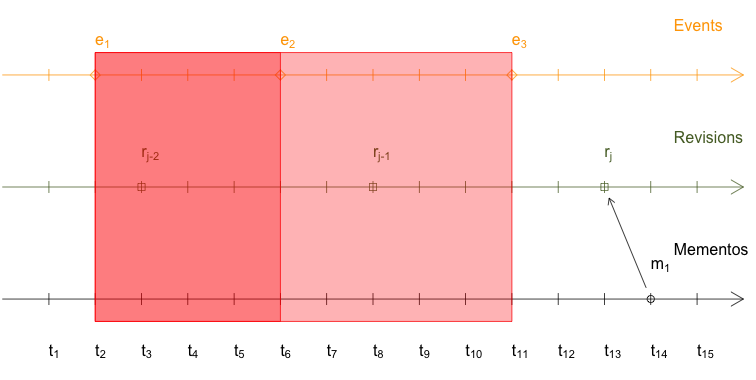}
	\caption{Example of \emph{pre-archive spoiler areas} (shown in light red) created using the mindist heuristic; the overlap of the spoiler areas for episodes $e_3$ and $e_2$ is shown in darker red.}	
	\label{fig:pre-archive-spoiler-area-definition}
\end{figure}

The set of datetimes where the user is directed to a spoiler, even though they chose a datetime prior to the episode they are avoiding, and where the web archive has not yet started archiving the resource, is referred to as a \textbf{pre-archive spoiler area}.  Figure \ref{fig:pre-archive-spoiler-area-definition} shows two pre-archive spoiler areas.  This spoiler area is created if the user tries to select a datetime prior to episode $e_3@t_{11}$, but the mindist heuristic delivers them to $m_1@t_{14} \equiv r_j@t_{13}$, which is after $e_3@t_{11}$.  The user intended to avoid spoilers for episode $e_3$, but got them nonetheless because the archive's earliest memento is after the desired datetime.  

So, for a pre-archive spoiler area to exist, the following conditions must be present:
\begin{enumerate}
\item The TimeGate for the resource uses the mindist heuristic
\item We have access to all revisions of a given resource
\item The Memento-Datetimes times for all revisions of a resource are defined and known
\item Event $e$ must occur prior to the first memento recorded in the archive
\item Event $e$ must occur prior to revision $r_i$ corresponding to the first memento $m_1$ (i.e., $r_i \equiv m_1 \land t_e < t_{r_j}$)
\end{enumerate}

Given episodes $e_1$ to $e_i$, which occur just prior to the first archived revision $r_j \equiv m_1$, this gives us the definition of a pre-archive spoiler area for episode $e_i$ defined by function $\mathcal{S}_a$ over the interval $t_s$ and ending at finish datetime $t_f$ produced by Equation \eqref{eq:pre-archive-spoiler-area-relationship}.

\begin{align}
[t_s, t_f] = \mathcal{S}_a(e_i) = \left\{
	\begin{array}{l l l}
		(t_{e_1}, t_{e_i}) & \textrm{    if     }  & t_{e_i} < t_{r_j} \\
		& \; \; \; \land  r_j \equiv m_k \\
		(0, 0) & \textrm{otherwise} & 
	\end{array}
\right.
\label{eq:pre-archive-spoiler-area-relationship}
\end{align}

\begin{figure}[t]
    \includegraphics[width=0.48\textwidth]{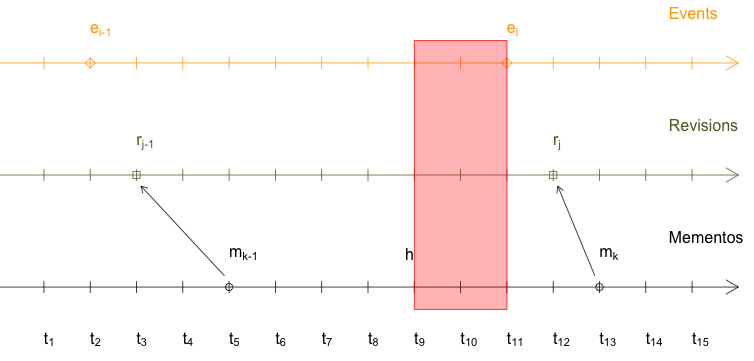}
    \caption{Example of a \emph{archive-extant spoiler area} (shown in light red) created by using the mindist heuristic, $h$ is the midpoint between $m_{k-1}$ and $m_k$} 
    \label{fig:archive-extant-spoiler-area-definition}
\end{figure}

Figure \ref{fig:archive-extant-spoiler-area-definition} shows an \textbf{archive-extant spoiler area}.  Let a user select a datetime prior to $e_i@t_{11}$.  To avoid spoilers, the user needs to be directed to memento $m_{k-1}$ corresponding to revision $r_{j-1}$.

Unfortunately, if the user selects a datetime in the area between $t_9$ and $e_i@t_{11}$, mindist will return memento $m_k@t_{13}$, even though they chose a datetime prior to $t_{11}$.  Memento $m_k@t_{13} \equiv r_j@t_{12}$, and $r_j$ exists \emph{after} the datetime $t_{11}$ that the user was trying to avoid. Because the user chose a datetime prior to the episode containing spoilers, but the user is redirected to a memento containing spoilers anyway.  

Why is this a spoiler area?  Remember that mindist finds the minimum distance between the time $t_a$ specified by the user and any given memento.  In Figure \ref{fig:archive-extant-spoiler-area-definition}, we have mementos $m_{k-1}@t_{5}$ and $m_{k}@t_{13}$.  We denote the midpoint between mementos as $h$ (for halfway). This means that any value $t_a$ such that $t_9 < t_a < t_{13}$ will produce memento $m_j$ and any value $t_a$ such that $t_a < t_9$ will produce memento $m_{j-1}$.

% The midpoint between $m_{k-1}@t_5$ and $m_k@t_{13}$ is $h@t_9$, calculated as shown in Equation (\ref{eq:midpoint-formula}).  

%\begin{align}
%t_h = \frac{t_i + t_j}{2}
%\label{eq:midpoint-formula}
%\end{align}

So, for a archive-extant spoiler area to exist, the following conditions must be present:
\begin{enumerate}
\item The TimeGate for the resource uses the mindist heuristic
\item We have access to all revisions of a given resource
\item The memento-datetimes times for all revisions of a resource are defined and known
\item Event $e$ must occur between the memento-datetimes of two consecutive mementos $m_{k-1}$ and $m_k$ (i.e.,  $t_{m_{k-1}} < t_e < t_{m_{k}}$)
\item Event $e$ must occur prior to revision $r_i$ corresponding to memento $m_j$ (i.e., $r_j \equiv m_k \land t_e < t_{r_j}$)
\item The midpoint $t_h$ between $m_{j-1}$ and $m_j$ must occur prior to event $e$: (i.e., $t_{m_{k-1}} < t_h < t_e < t_{m_k}$)
\end{enumerate}

Given consecutive mementos $m_{k-1}$ and $m_k$, the midpoint $t_h$ between them, and revision $r_j \equiv m_k$, this gives us the definition of an archive-extant spoiler area defined by function $\mathcal{S}_b$ over the interval beginning at start datetime $t_s$ and ending at finish datetime $t_f$ produced by Equation \eqref{eq:archive-extant-spoiler-area-relationship}.
\begin{align}
[t_s, t_f] = \mathcal{S}_b(e) = \left\{
	\begin{array}{l l l}
		(t_h, t_e) & \textrm{  if   } t_h < t_e < t_{r_i} \\
		& \; \; \; \textrm{ } \land \textrm{ } r_j \equiv m_k  \textrm{ } \land \\
		& \; \; \; \; t_h = \frac{t_{m_{k-1}} + t_{m_k}}{2} \\
		(0, 0) & \textrm{otherwise} 
	\end{array}
\right.
\label{eq:archive-extant-spoiler-area-relationship}
\end{align}

%The spoilers we have been discussing are not for single events, but episodes of a larger story.  Our assumption is that \emph{single} episode stories are revealed in an instant, on a specific date, and anyone trying to find specific, deep information about them can not be served by our method if they wish to avoid spoilers.  

So, how does one handle multiple episodes?  What does that mean for our spoiler areas?  For a given resource, using mindist, what is the chance of attempting web time travel with Memento and getting a spoiler?

\begin{figure}[t]
    \includegraphics[width=0.48\textwidth]{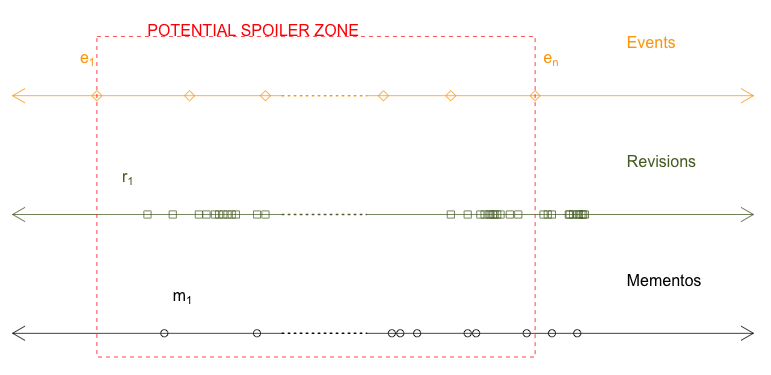}
    \caption{Example of a potential spoiler zone, stretching from $t_{e_1}$ to $t_{e_n}$}    
    \label{fig:potential-spoiler-zone}
\end{figure}

First we define a \textbf{potential spoiler zone} across the length of the series we are looking at.  The start datetime of the potential spoiler zone is $t_{e_1}$, the datetime of the first episode.  The end datetime of our potential spoiler zone is $t_{e_n}$, the datetime of the last (or latest) episode.  We assume that a user searching for datetimes prior to the first event $e_1$ should get no spoilers, so that is the lower bound.  We also assume that no additional spoilers can be revealed after the last event $e_n$.  This provides a single area in which we can determine the probability of getting a spoiler for a single episode in the series.  Figure \ref{fig:potential-spoiler-zone} shows an example of such a zone.

\begin{figure}[t]
    \includegraphics[width=0.48\textwidth]{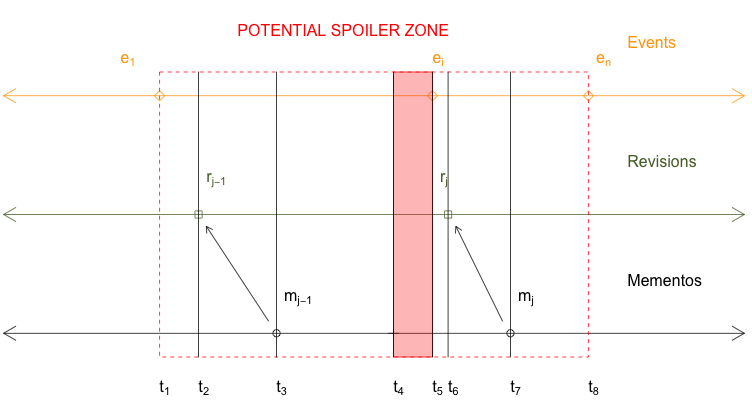}
    \caption{Example of a spoiler area (light red area) for episode $e_i$ inside potential spoiler zone (dotted red rectangle), stretching from $t_{e_1}$ to $t_{e_n}$}    
    \label{fig:spoiler-area-in-potential-spoiler-zone}
\end{figure}

Figure \ref{fig:spoiler-area-in-potential-spoiler-zone} shows a spoiler area ($t_4$ to $t_5$) inside a potential spoiler zone ($t_{e_1}$ to $t_{e_n}$).  Consider randomly choosing a desired datetime within this zone.  What is the probability of landing inside the spoiler area for given episode $e$?

Probability is defined as the number of times something can occur divided by the total number of outcomes \cite{Yaspan1968}.  The smallest unit of datetime on the web is the second.  We cannot gain more precision over time due to the fact that HTTP headers (and hence Memento-Datetimes) use the second as the smallest unit.  Consider iterating through every second between $e_1$ and $e_n$, incrementing the value of counter $s$ for each second that falls within a spoiler area.  If we let $c$ be the number of seconds between $e_1$ and $e_n$, then the probability of encountering a spoiler is shown by equation \eqref{eq:probability-of-spoiler}.

\begin{align}
Pr(spoiler) = \frac{s}{c}
\label{eq:probability-of-spoiler}
\end{align}

Once we have determined the probability of encountering a spoiler for a resource within the Internet Archive, we can then use that probability to compare that resource to others.  In this way we can determine how safe a given URI is for users who want to avoid spoilers using the Wayback Machine or a Memento TimeGate that uses the mindist heuristic.

\section{Measuring Spoiler Probability in Popular Wikis}

\begin{table*}[htbp]
\caption{Fan wikis used in the spoiler areas experiment} 
\label{tab:wikis-used}

\small
\begin{center}
\begin{tabular}{| l | l | l | l | l | l |}
\hline
\textbf{Television Show \newline (Network)} & \textbf{Wiki URI}                      & \textbf{\# of Pages} & \textbf{$t_{r_1}$} & \textbf{$t_{e_1}$} & \textbf{\%  of pages in} \\
& & & & & \textbf{Internet} \\
& & & & & \textbf{Archive} \\
\hline
the Big Bang Theory (CBS)          & \url{bigbangtheory.wikia.com}         & 1120                     & 2007-12-14            & 2007-09-24   &  68.8\%                \\
\hline
Boardwalk Empire (HBO)             & \url{boardwalkempire.wikia.com}       & 2091                     & 2010-03-18            & 2010-08-23   &  80.6\%               \\
\hline
Breaking Bad (A\&E)                & \url{breakingbad.wikia.com}           & 998                      & 2009-04-27            & 2008-01-20   &  76.0\%                \\
\hline
Continuum (Showcase)               & \url{continuum.wikia.com}             & 258                      & 2012-11-13            & 2012-05-27   &  86.8\%                \\
\hline
Downton Abbey (BBC)                & \url{downtonabbey.wikia.com}          & 784                      & 2010-10-04            & 2010-09-26   &  53.1\%               \\
\hline
Game of Thrones (HBO)              & \url{gameofthrones.wikia.com}         & 3144                     & 2010-06-24            & 2011-04-17   &  75.8\%                \\
\hline
Grimm (NBC)                        & \url{grimm.wikia.com}                 & 1581                     & 2010-04-14            & 2011-10-28   &  57.5\%               \\
\hline
House of Cards (Netflix)           & \url{house-of-cards.wikia.com}        & 251                      & 2013-01-11            & 2013-02-01   &  97.2\%                \\
\hline
How I Met Your Mother (CBS)        & \url{how-i-met-your-mother.wikia.com} & 1709                     & 2008-07-21            & 2005-09-19   &  58.7\%                \\
\hline
Lost (ABC)                         & \url{lostpedia.wikia.com}             & 18790                    & 2005-09-22            & 2004-09-22    &  39.1\%               \\
\hline
Mad Men (AMC)                      & \url{madmen.wikia.com}                & 652                      & 2009-07-25            & 2007-06-03    &  85.0\%               \\
\hline
NCIS (CBS)                         & \url{ncis.wikia.com}                  & 5345                     & 2006-09-25            & 2003-09-23    &  93.2\%               \\
\hline
Once Upon A Time (ABC)             & \url{onceuponatime.wikia.com}         & 1470                     & 2011-08-09            & 2011-10-23    &  79.9\%               \\
\hline
Scandal (ABC)                      & \url{scandal.wikia.com}               & 331                      & 2011-06-07            & 2012-04-05    &   82.8\%              \\
\hline
True Blood (HBO)                   & \url{trueblood.wikia.com}             & 1838                     & 2008-10-06            & 2008-09-07    &   74.1\%              \\
\hline
White Collar (USA)                 & \url{whitecollar.wikia.com}           & 506                      & 2009-10-30            & 2009-10-23    &   79.1\%             \\
\hline
\end{tabular}
\end{center}

\caption{Spoiler probabilities for most popular pages within each fan wiki}
\label{tab:pop-pages}
\begin{center}
\begin{tabular}{|l|l|l|l|l|l|}
\hline
\textbf{Wiki}         & \textbf{Page Name} & \textbf{Probability} & \textbf{\# of} & \textbf{\# of} & 
\textbf{\# of} \\ 
& & \textbf{of Spoiler} & \textbf{Spoiler Areas} & \textbf{Revisions} & \textbf{Mementos} \\
\hline
bigbangtheory         & Sheldon Cooper     & 0.31                            & 69                           & 1958                     & 30                      \\ \hline
boardwalkempire       & Nucky Thompson     & 0.15                            & 31                           & 290                      & 15                      \\ \hline
breakingbad           & Walter White       & 0.43                            & 40                           & 882                      & 20                      \\ \hline
continuum             & Keira Cameron      & 0.54                            & 21                           & 104                      & 5                       \\ \hline
downtonabbey          & Sybil Branson      & 0.42                            & 23                           & 580                      & 3                       \\ \hline
gameofthrones         & Daenerys Targaryen & 0.16                            & 24                           & 768                      & 29                      \\ \hline
grimm                 & Nick Burkhardt     & 0.39                            & 30                           & 795                      & 5                       \\ \hline
house-of-cards        & Frank Underwood    & 0.0                             & 13                          & 380                      & 3                       \\ \hline
how-i-met-your-mother & Barney Stinson     & 0.55                            & 120                          & 588                      & 13                      \\ \hline
lostpedia             & Kate Austen        & 0.67                            & 94                           & 3531                     & 27                      \\ \hline
madmen                & Mad Men Wiki       & 0.22                            & 36                           & 250                      & 85                      \\ \hline
ncis                  & Abigail Sciuto     & 0.67                            & 182                          & 404                      & 11                      \\ \hline
onceuponatime         & Emma Swan          & 0.36                            & 34                           & 1210                     & 11                      \\ \hline
scandal               & Main Page          & 0.60                            & 31                           & 250                      & 14                      \\ \hline
trueblood             & Eric Northman      & 0.28                            & 47                           & 931                      & 14                      \\ \hline
whitecollar           & Neal Caffrey       & 0.29                            & 38                           & 199                      & 8                       \\ \hline
\end{tabular}
\end{center}

\caption{Statistics for each fan wiki}
\label{tab:spoiler-statistics}
\begin{center}
\begin{tabular}{|l||l|l|l||l|l|l||l|l|l|}
\hline
\multirow{3}{*}{\textbf{Wiki}} & \multicolumn{3}{|c|}{\textbf{Probability of Spoiler}} & \multicolumn{3}{|c|}{\textbf{Revisions/Day}} & \multicolumn{3}{|c|}{\textbf{Mementos/Day}} \\
\cline{2-10}
& \textbf{Mean} & \textbf{std dev} & \textbf{Rel Err} & \textbf{Mean} & \textbf{std dev} & \textbf{Rel Err} & \textbf{Mean} & \textbf{std dev} & \textbf{Rel Err}\\ 
\hline
bigbangtheory  &  0.667  &  0.160  &  0.0116  &  0.0506  &  0.0668  &  0.0639  &  0.0033  &  0.0034  &  0.0488 \\
\hline
boardwalkempire  &  0.417  &  0.170  &  0.0160  &  0.0102  &  0.0185  &  0.0718  &  0.0022  &  0.0026  &  0.0452 \\
\hline
breakingbad  &  0.746  &  0.205  &  0.0127  &  0.0185  &  0.0351  &  0.0872  &  0.0032  &  0.0032  &  0.0459 \\
\hline
continuum  &  0.394  &  0.177  &  0.0471  &  0.0317  &  0.0250  &  0.0829  &  0.0051  &  0.0023  &  0.0479 \\
\hline
downtonabbey  &  0.585  &  0.174  &  0.0196  &  0.0374  &  0.0636  &  0.1124  &  0.0020  &  0.0013  &  0.0419 \\
\hline
gameofthrones  &  0.473  &  0.248  &  0.0122  &  0.0425  &  0.0652  &  0.0356  &  0.0041  &  0.0049  &  0.0279 \\
\hline
grimm  &  0.479  &  0.175  &  0.0201  &  0.0700  &  0.0857  &  0.0672  &  0.0027  &  0.0015  &  0.0305 \\
\hline
house-of-cards  &  0.006  &  0.035  &  0.6705  &  0.0772  &  0.1364  &  0.2082  &  0.0075  &  0.0044  &  0.0687 \\
\hline
how-i-met-your-mother  &  0.741  &  0.100  &  0.0046  &  0.0163  &  0.0220  &  0.0463  &  0.0014  &  0.0010  &  0.0263 \\
\hline
lostpedia  &  0.768  &  0.163  &  0.0027  &  0.0391  &  0.1083  &  0.0348  &  0.0040  &  0.0055  &  0.0173 \\
\hline
madmen  &  0.530  &  0.144  &  0.0133  &  0.0049  &  0.0076  &  0.0764  &  0.0014  &  0.0021  &  0.0755 \\
\hline
ncis  &  0.818  &  0.107  &  0.0041  &  0.0073  &  0.0097  &  0.0413  &  0.0009  &  0.0008  &  0.0279 \\
\hline
onceuponatime  &  0.516  &  0.163  &  0.0132  &  0.1271  &  0.1327  &  0.0437  &  0.0037  &  0.0025  &  0.0281 \\
\hline
scandal  &  0.591  &  0.165  &  0.0269  &  0.0418  &  0.0484  &  0.1120  &  0.0030  &  0.0019  &  0.0608 \\
\hline
trueblood  &  0.517  &  0.162  &  0.0106  &  0.0210  &  0.0410  &  0.0658  &  0.0016  &  0.0016  &  0.0345 \\
\hline
whitecollar  &  0.390  &  0.250  &  0.0500  &  0.0117  &  0.0147  &  0.0986  &  0.0019  &  0.0015  &  0.0609 \\
\hline
\textbf{Overall} & \textbf{ 0.659 } & \textbf{ 0.226 } & \textbf{ 0.0029 } & \textbf{ 0.0362 } & \textbf{ 0.0871 } & \textbf{ 0.0200 } & \textbf{ 0.0032 } & \textbf{ 0.0044 } & \textbf{ 0.0114 } \\
\hline
\end{tabular}
\end{center}
\end{table*}

We selected 16 fan wikis based on television shows for our experiment. Table \ref{tab:wikis-used} shows some of details for each fan wiki.  Each television show selected has had at least two seasons and a currently active wiki.  \emph{House of Cards} was chosen because an entire season is released on Netflix in a single day, making it different from networked television shows.  \emph{Lost} was chosen because its wiki, \emph{Lostpedia}, has undergone academic study \cite{Mittell2009}, and is the oldest and largest fan wiki under consideration.  We used a process, simplified in Algorithm \ref{alg:spoiler-areas}, to process each wiki and identify the spoiler areas created by mindist.  Episode dates were supplied by epguides.com.

\begin{algorithm}[t]
\caption{Algorithm for spoiler probability experiment}
\label{alg:spoiler-areas}
\begin{codebox}
\Procname{$\proc{FindSpoilerAreasInWikis}(episodeList, wikiURI)$}
\li $episodeTimes \gets \proc{getEpisodeTimes(episodeList)}$
\li $wikiTitles \gets \proc{getPageTitles}(wikiURI)$
\li \For each $title \in wikiTitles$
	\Do
\li	$wikidump \gets \proc{fetchXMLdump}(title, wikiURI)$
\li	$revisions \gets \proc{extractRevisionTimes}(wikidump)$
\li	$timemapURI \gets \proc{makeTMURI}(wikiURI, title)$
\li	$timemap \gets \proc{fetchTimeMap}(timemapURI)$
\li	$mementos \gets \proc{extractMementoTimes}(timemap)$
\li	$mementoRevisionMap \gets$
\\ \quad \quad \quad $\proc{mapRevsToMems}(revisions, mementos)$
\li	\For each $episode \in episodeTimes$
		\Do
\li 		$paSpoilerArea \gets$
\\ \quad \quad \quad \quad \quad $S_a(episode, mementoRevisionMap$
\li		$aeSpoilerArea \gets$
\\ \quad \quad \quad \quad \quad $S_b(episode, mementoRevisionMap)$
\li		$spoilerAreaList.append(paSpoilerArea)$
\li		$spoilerAreaList.append(aeSpoilerArea)$
		\End
\li	$\proc{mapPageToSpoilers}($
\\ \quad \quad \quad $wikipageSpoilerMap, title, spoilerAreaList)$
\End
\li \Return $wikipageSpoilerMap$
\end{codebox}
\end{algorithm}

%The articles of each wiki were analyzed and processed to find spoiler areas using the following process:
%\begin{enumerate}
%\item Fetch the list of titles for all articles on the wiki using the MediaWiki web API
%\item Iterate through this list, fetching the full-history XML dumps of each article using MediaWiki's export functionality
%\item Process the XML dumps to acquire the datetimes of the revisions for each article using the Python libraries and programs provided by Wikimedia Utilities
%\item Construct TimeMap URIs based on the base URI of the wiki, the article name, and the base TimeMap URI at the Internet Archive
%\item Fetch the TimeMaps from these URIs using the \texttt{wget} program, saving them to filenames generated based on the MD5 version of the original URI to avoid any filename encoding issues with further processing
%\item Use Unix utilities, such as \texttt{sed}, on the TimeMaps to acquire the datetimes for the mementos for each article in epoch time format
%\item Calculate the midpoints between the mementos for each article in epoch time format
%\item Map mementos to the revisions they archived
%\item Fetch the list of episodes and episode datetimes for the television show and convert them to epoch time
%\item Using the equations \eqref{eq:pre-archive-spoiler-area-relationship} and \eqref{eq:archive-extant-spoiler-area-relationship}, calculate the pre-archive and archive-extant spoiler areas for each episode
%\end{enumerate}

Utilizing this method, we computed additional statistics based on the revisions, mementos, the memento-revision mapping, and the spoiler areas.

Out of the 40,868 wiki pages processed for this experiment, we discovered that many of them were wiki redirects.  Redirects are used to deal with articles that can be referred to by multiple names.  Sometimes wiki editors may not know the real name of an introduced character until much later, and will use a redirect from the old name to the new.  Sometimes wiki editors will create pages not knowing that one already exists, leaving future editors to create a redirect now that they know that a new page title was desired.  Because of the number of redirects that contained only a single revision and only a single memento, we removed the redirects from consideration for calculation of spoiler areas and other statistics.  This removed 16,394 pages from consideration, leaving us with 24,474 pages to process.

The wiki XML exports were downloaded at a different time than the TimeMaps for those wiki pages.  To overcome this inconsistency, any mementos in TimeMaps that existed after the wiki page was downloaded were discarded.

Of the 24,474 pages processed, only 15,119 pages actually had TimeMaps at the Internet Archive at the time the wiki exports were extracted.  This means that roughly 38\% of the pages under consideration were not available in the Internet Archive.

\begin{figure}[htbp]
	\includegraphics[width=0.5\textwidth]{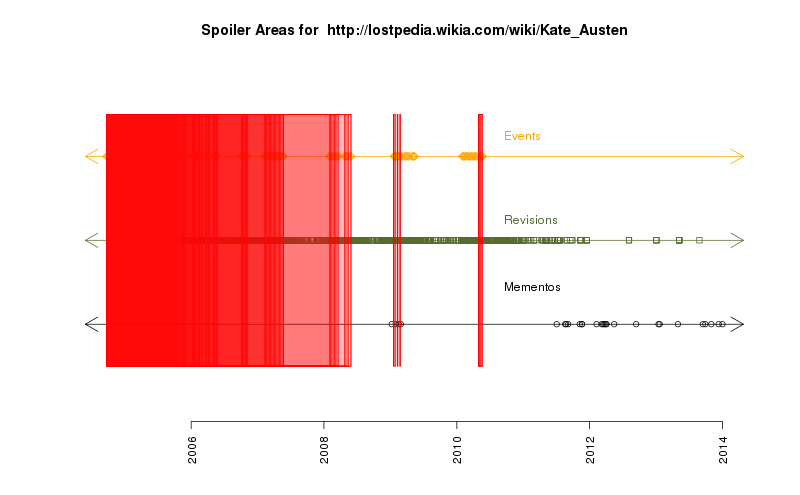}
	\caption{Spoiler areas for the most popular page (3,531 revisions) in our data set}
	\label{fig:spoiler-areas-Kate-Austen}

	\includegraphics[width=0.5\textwidth]{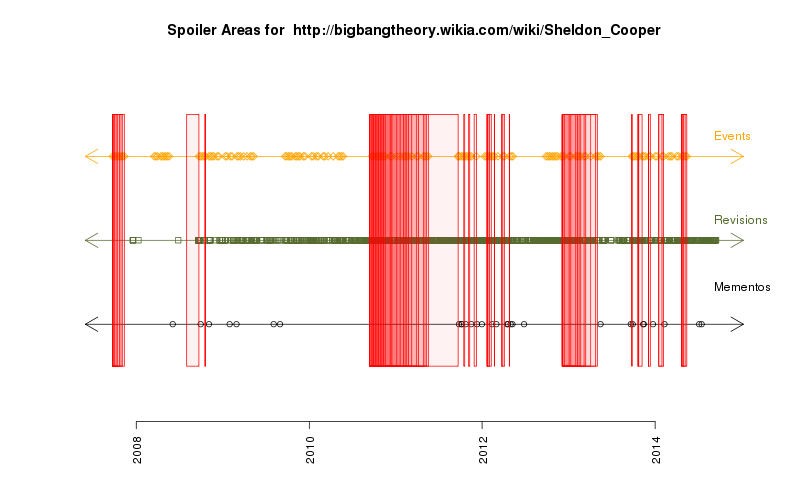}
	\caption{Spoiler areas for the most popular page (1,958 revisions) in \emph{the Big Bang Theory Wiki}}
	\label{fig:spoiler-areas-Sheldon-Cooper}
	
	\includegraphics[width=0.5\textwidth]{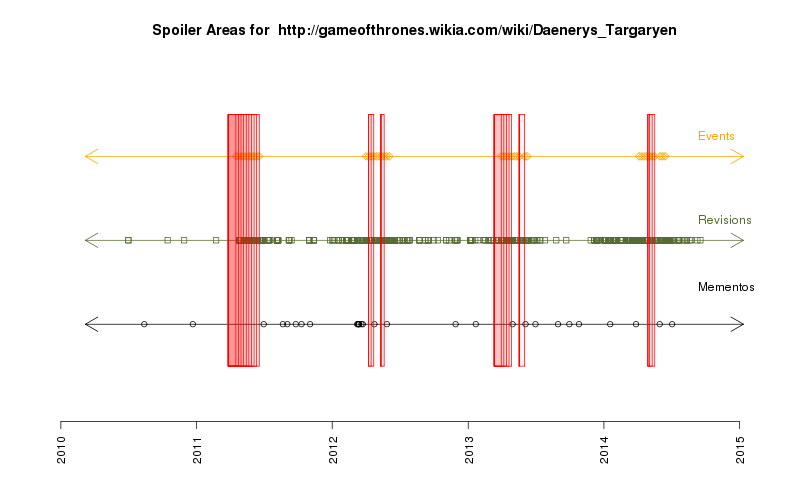}
	\caption{Spoiler areas for the most popular page (768 revisions) in the \emph{Game of Thrones Wiki}}	
	\label{fig:spoiler-areas-Daenerys-Targaryen}
	
\end{figure}

\begin{figure*}[ht]
	\begin{minipage}[b]{0.47\linewidth}
		\includegraphics[width=1\textwidth]{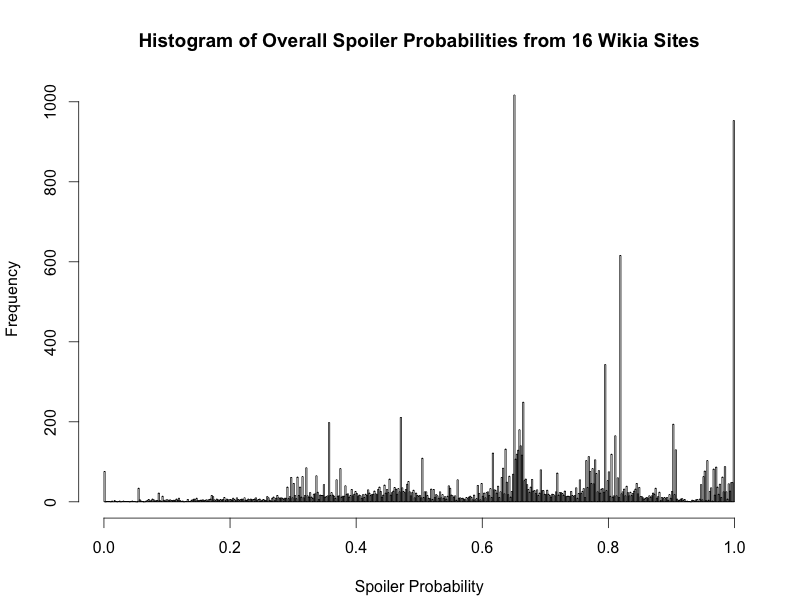}
		\caption{Histogram of spoiler probabilities for all 16 wiki sites}
		\label{fig:hist-all.png}
	\end{minipage}%
	\hfill{}
	\begin{minipage}[b]{0.47\linewidth}
		\includegraphics[width=1\textwidth]{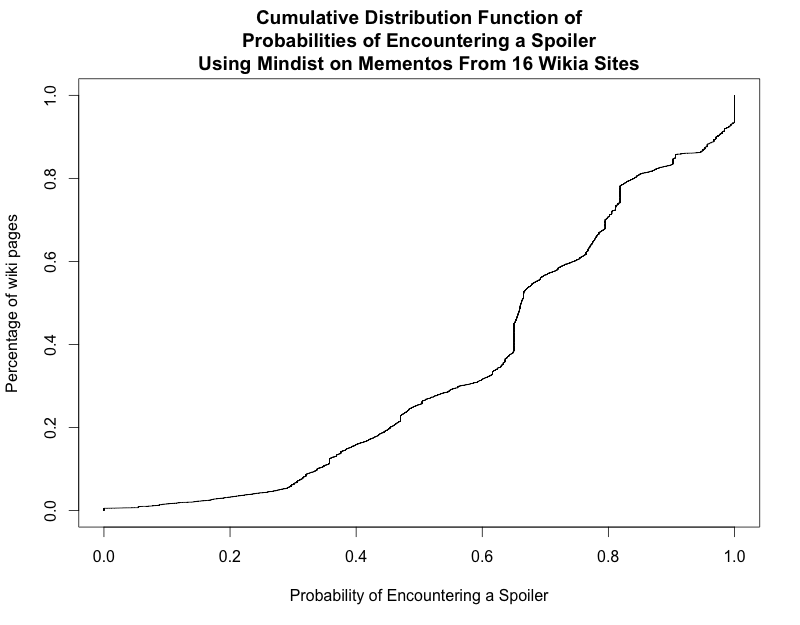}
		\caption{Graph of the cumulative distribution function of spoiler probabilities for all 16 wiki sites}
		\label{fig:hist-spoilers-cdf-all.png}
	\end{minipage}
	
	\begin{minipage}[b]{0.47\linewidth}
		\includegraphics[width=1\textwidth]{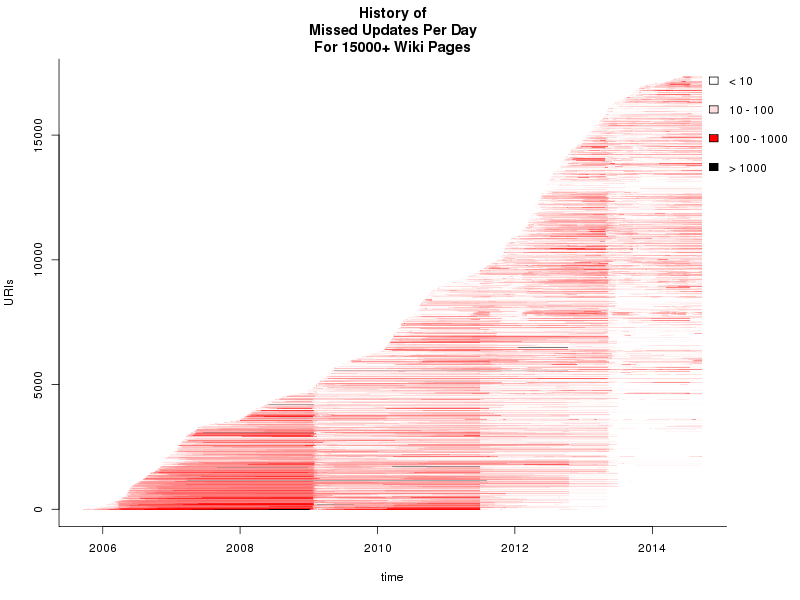}
		\caption{Visualization of missed updates; darker colors represent more missed updates}		
		\label{fig:mu-all-history}
	\end{minipage}%
	\hfill{}
	\begin{minipage}[b]{0.47\linewidth}
		\includegraphics[width=1\textwidth]{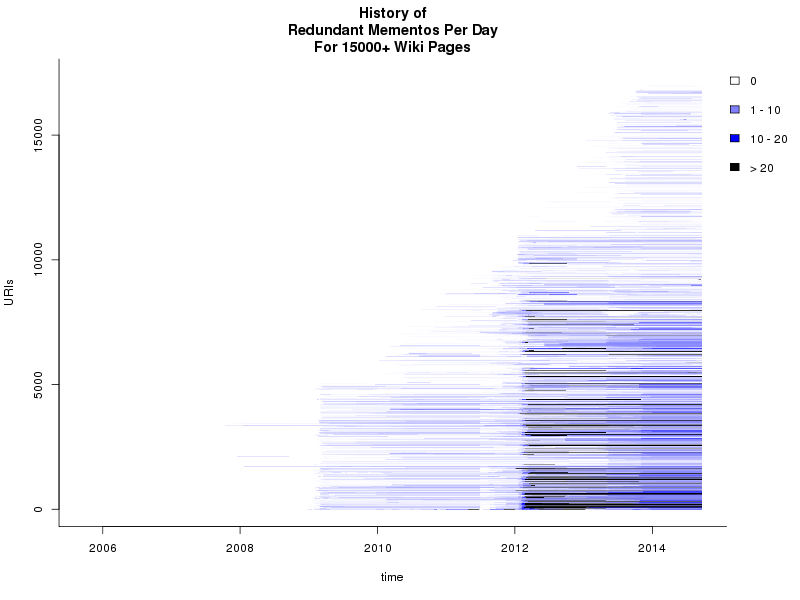}
		\caption{Visualization of redundant mementos; darker colors represent more redundant mementos}		
		\label{fig:rm-all-history}
	\end{minipage}
	
\end{figure*}

%\footnotetext{\url{http://lostpedia.wikia.com/wiki/Kate_Austen}}

Figure \ref{fig:spoiler-areas-Kate-Austen} shows our spoiler area graph for the page with the most revisions in our entire dataset, a page from \emph{Lostpedia} about the character Kate Austen.  Each spoiler area is shown in red using an alpha channel that gives it some degree of transparency.  When these transparent red areas stack up, of course the red gets darker, so we cannot reliably see all of the pre-archive spoiler areas that exist prior to the first memento.  The probability of encountering a spoiler for Kate's page is $67\%$, calculated by Equation \eqref{eq:probability-of-spoiler}.  Because this page only has a few mementos around 2009 and then a long break for the Internet Archive until 2011, there are a few archive-extant spoiler areas, also shown in red, both around the 2009 mark.  We also see some archive-extant spoiler areas, and also after the memento halfway mark in 2010.

%\footnotetext{\url{http://bigbangtheory.wikia.com/wiki/Sheldon_Cooper}}

Figure \ref{fig:spoiler-areas-Sheldon-Cooper} shows spoiler areas for the page about the \emph{Big Bang Theory} character Sheldon Cooper.  The Internet Archive is more aggressive at archiving in 2008 than it was during the run of the show \emph{Lost} (starting 2004), so there are only 8 pre-archive spoiler areas for this page, compared to Kate's 86.  There are, however, 61 archive-extant spoiler areas, compared with Kate's 8.  Sheldon's page has a spoiler probability of only $31\%$.  We can see the clusters of points indicating each episode on the events timeline.  Because television show seasons occur during portions of the year, we can see the seasons, and partial seasons, for \emph{Big Bang Theory} on the top.  Even though Sheldon's page contains quite a few spoiler areas after the second season, there appears to be a block of time before the third season where one is safe to browse this page and avoid spoilers.

Figure \ref{fig:spoiler-areas-Daenerys-Targaryen} provides another example of a more current show, using a page from the \emph{Game of Thrones Wiki}.

Table \ref{tab:pop-pages} contains statistics for the most popular page in each of the wikis that we have surveyed, where popularity is determined by the number of page revisions generated.  Seeing as these wikis are authored by fans, readers familiar with many of these television shows will not be surprised that most of the popular pages are main characters.  The table also lists the number of spoiler areas, revisions, and mementos, showing how there is not a simple relationship between these values that indicate the probability of encountering a spoiler.

Of particular interest is the television show, \emph{House of Cards}.  Because it releases an entire season of episodes at one time, our model breaks down.  We count $13$ pre-archive spoiler areas for the first season, and then no archive-extant spoiler areas.  The pre-archive spoiler areas have no size due to the fact that all of them begin and end at the same time.  This leads to a $0\%$ chance of encountering a spoiler in this wiki, seeing as each season is released like a 13-hour movie rather than on a weekly basis. In this case, time is not able to differentiate between individual episodes because $t_{e_1} = t_{e_2} = \ldots t_{e_{13}}$.  It requires a new dimension in order to order otherwise simultaneous events.  A different situation exists with another Netflix series, \emph{Arrested Development}, in which all episodes for a season are released at once, but the episodes do not need to be viewed in any particular order, making it difficult to identify when spoilers would occur.

Table \ref{tab:spoiler-statistics} shows the statistics for each fan wiki.  We see a mean overall spoiler probability of 66\%.  We also see that the number of mementos per day is an order of magnitude smaller than the number of revisions per day.

Figure \ref{fig:hist-all.png} shows the probability distribution of encountering spoilers in these wiki pages.  Figure \ref{fig:hist-spoilers-cdf-all.png} shows a cumulative distribution function of spoiler probabilities for all wikis within the data set.  Here we see that the spoiler probability exists, in some form, for most of the pages.

Figure \ref{fig:mu-all-history} shows the number of missed updates encountered for each datetime over the history of all pages in the wiki.  The Y-axis represents each URI in the data set.  The X-axis is time. Lighter colors indicate fewer missed updates on that day.  Of interest are the vertical lines seen throughout the visualization.  The datetimes for these lines correspond to changes in policy at the Internet Archive.  In 2009 and in late 2011, the Internet Archive reduced its quarantine period for archiving of new pages.  In October of 2013, the Internet Archive published the \emph{Save Page Now} feature  \cite{Rossi2013}, leading to fewer missed updates after that point.

Figure \ref{fig:rm-all-history} shows the number of redundant mementos created for each datetime over the history of all pages in the wiki.  Just as with Figure \ref{fig:mu-all-history}, the Y-axis represents each URI and the X-axis is time.  As expected, the number of redundant mementos increases as the Internet Archive becomes more aggressive about archiving web pages.

\section{Measuring Na{\"i}ve Spoilers in \newline Wayback Machine Logs}

Research has already been done by Ainsworth in how much drift exists within the web archive \cite{Ainsworth2013}.  That study indicates that the Wayback Machine uses a \textbf{sliding target policy}.   This means that each request is in some way based on the datetime of the last request, resulting in a user ending up in a much different datetime than they had originally started.  The Wayback Machine still uses the mindist heuristic to determine which memento to deliver to a user, but it changes the desired datetime $t_a$ based on the datetime of the memento from the last request.

Contrary to this, Memento uses a \textbf{sticky target policy}, allowing a user to fix the datetime $t_a$ throughout their browsing session.  While the sparsity of the archives introduces some small drift with the sticky target policy, it is constrained by the datetime remaining constant in each request.  That drift is introduced only by the mindist heuristic rather than the sliding behavior of the Wayback Machine.

We are concerned about whether or not the user ended up in the future of where they intended.  We want to know if they encountered a spoiler when using the Wayback Machine.  We conducted a studying using anonymized Wayback Machine logs spanning January 1, 2011 through March 10, 2011 and August 1, 2011 through March 26, 2012.

%\begin{figure}[htbp]
%\caption{URI-M pattern for the Wayback Machine and Internet Archive}
%\includegraphics[width=0.46\textwidth]{images/wayback-machine-uri-pattern.png}
%\label{fig:wayback-machine-url-pattern}
%\end{figure}

The logs from the Wayback Machine are in Apache common log format.  Using the referrer for each request, we can track where the user came from and determine where they ended up.  Fortunately for us, we can infer the desired datetime (referred to as $t_a$ previously) and the memento-datetime from the URIs themselves.  The Internet Archive allows access to all mementos using a standard URI format and the datetime is embedded in the URI.  For the URI visited by the user, this datetime indicates the memento-datetime.  For the referrer URI, this datetime indicates their desired datetime.
%Figure \ref{fig:wayback-machine-url-pattern} shows the URI pattern used by the Internet Archive and the Wayback Machine to identify mementos.  

\begin{algorithm}[t]
\caption{Algorithm for Detecting spoilers in Internet Archive Logs}
\label{alg:log-spoiled}
\begin{codebox}
\Procname{$\proc{FindSpoilersInLogFile}(logfile)$}
\li \For each  $visitor ID$, $visited URI$, $referrer \in logfile$
 	\Do
%	$(visitor ID, visited URI, referrer) \gets \proc{getValues}(line)$
\li	$t_m \gets \proc{getDate}(visited URI)$
\li	$t_a \gets \proc{getDate}(referrer)$
\li $wikidump \gets \proc{fetchXMLdump}(title, wikiURI)$
\li $revisions \gets \proc{extractRevisionTimes}(wikidump)$
\li $t_r \gets \proc{getRevMatchingMemento}(t_m, revisions)$
\li	$spoiler \gets INDETERMINATE$
\li	\If $rev$ is not $NULL$
	\Then
\li		$spoiler \gets (t_a < t_r)$
	\End			
\li	\proc{print}($visitor ID$ + " , " + $spoiler$)
	\End
\end{codebox}
\end{algorithm}

Why do we say that we can \emph{infer} the desired datetime?  Without interviewing the visitors to the Wayback Machine, it is impossible to determine intent.  The fact that the logs are anonymized makes this completely impossible.  We are making the assumption that some of the users receiving these responses intended to receive responses on the date that they started at, not the date delivered by the drift caused by the mindist heuristic.

From these logs we can determine the inferred desired datetime from the referrer URI and the memento-datetime from the visited URI.  Using this information, we can download the wiki exports, as in the previous experiment, and determine if the page revision recorded by the web archive exists in the future of the desired datetime.

All requests for archived pages from wikia.com were extracted from the logs, resulting in 1,180,759 requests.  Of those requests, we removed all requests for images, JavaScript, style sheets, supporting wiki pages (such as Template, Category, and Special pages), and advertisements.  This left us with 62,227 requests to review.

For those remaining wikia.com pages, we downloaded the wiki export files, as done in the previous experiment, mapped the visited URI to the request that it had archived, and compared the datetime of that revision with the inferred desired datetime.  We use $t_a$ to represent the inferred desired datetime, and $t_r$ to represent the datetime of the wiki revision matching the visited URI in the Wayback Machine.

%Using this information, we can see how many requests end up in the future, meaning that those visitors are being redirected to spoilers via the Wayback Machine.

%In our discussions, we shall use $m_v$ as the visited URI-M and $m_{ref}$ as the referrer URI-M.  So, using this notation, $t_{m_{ref}}$ can be thought as equivalent to the desired datetime $t_a$ used earlier because it is our inferred desired datetime.

%We shall use $r_v$ as the URI-M of the wiki revision corresponding to the visited URI and $m_{ref}$ as the referrer URI-M.  So, using this notation, $t_{m_{ref}}$ can be thought as equivalent to the desired datetime $t_a$ used earlier because it is our inferred desired datetime.

%Each response can be split into four categories in terms of spoilers:  (1) \textbf{spoiler} - $t_{m_v} > t_{m_{ref}}$; (2) \textbf{safe} - $t_{m_v} \le t_{m_{ref}}$; (3) \textbf{indeterminate} - either the datetime for the visited URI-M or the referrer was not able to be determined, likely because they were not URI-Ms, but instead were Wayback Machine embedded images, stylesheets, or other web site pages; (4) \textbf{error} - something happened with the processing, typically related to corrupted log entries or log files; any log generating an error was removed from the pool of data.

Each response can be split into three categories in terms of spoilers:  (1) \textbf{spoiler} - $t_{a} < t_{r}$; (2) \textbf{safe} - $t_{a} \ge t_{r}$; (3) \textbf{indeterminate} - either the datetime for the revision or the referrer was not able to be determined, likely because the article or whole wiki was moved or no longer exists, or because of 503 HTTP status codes due to the size of the export file.

This process, shown in Algorithm \ref{alg:log-spoiled} determines how many requests are either spoiler, safe, or indeterminate for each log file. Indeterminate entries make up the bulk of the data collected, but offer no meaningful insight into the spoiler problem, and are thus discarded.  From this study we found that roughly 19\% of these requests to the Wayback Machine result in spoilers.

%\begin{table}[t]
%\caption{Results of scanning the logs for spoilers if we remove indeterminate results}
%\label{tab:wayback-mean-results}
%\small
%\begin{center}
%\begin{tabular}{l | l | l | l}
%\hline
%\textbf{Specific} & \textbf{Mean} & \textbf{Mean} & \textbf{\% of} \\
%\textbf{Subset} & \textbf{spoilers}  & \textbf{safe} & \textbf{spoiled} \\
%\textbf{} & \textbf{per} & \textbf{per} & \textbf{per} \\
%& \textbf{per day} & \textbf{day} & \textbf{day} \\
%\hline
%*.wikipedia.org & 9,691 & 27,065 & 26\% \\
%\hline
%*.wikia.com & 7,867 & 23,789 & 25\% \\
%\hline
%\end{tabular}
%\end{center}
%\end{table}

\section{Conclusions}
We have introduced the notion of different heuristics for use with Memento TimeGates.  We have shown that the mindist heuristic, while useful for sparse archives, is not reliably effective for users trying to avoid spoilers with Memento.  We have also proposed minpast as a superior choice for wikis, who have access to every revision.

We have shown that roughly 38\% of the pages under consideration were not available in the Internet Archive.  We also found that, for the wiki sites under consideration, there is a mean 66\% probability that one will end up with a spoiler if they use TimeGates supporting the mindist heuristic. Also, from our sample logs from the Wayback Machine, 19\% of requests to wikia.com end in spoilers. This presents a problem for episodic fiction fans trying to use the Wayback Machine, or the Internet Archive through Memento, to avoid spoilers.  This further demonstrates that using Memento directly on wikis, using minpast, is better for avoiding spoilers.

\section{Acknowledgments}
This work made possible in part by a grant from the Andrew Mellon Foundation.  We are grateful to the Internet Archive for their continued support of Memento.  Memento is a joint project between the Los Alamos National Laboratory Research Library and Old Dominion University.

\bibliographystyle{abbrv}
\bibliography{techreport}  % sigproc.bib is the name of the Bibliography in this case
% ACM needs 'a single self-contained file'!
%\balancecolumns

\onecolumn
\appendix

\begin{figure*}[!htp]
    \centering
	\includegraphics[width=\bigImageSize]{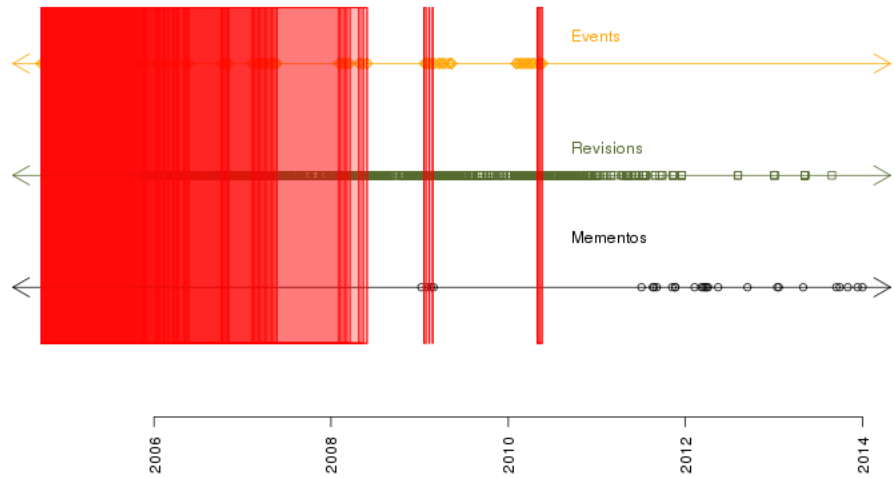}
	\caption{Spoiler areas for the most popular page in \emph{Lostpedia}\protect\footnotemark}
	\label{fig:spoiler-areas-Kate-Austen-2}
\end{figure*}
\footnotetext{\url{http://lostpedia.wikia.com/wiki/Kate_Austen}}

\begin{figure*}[!htp]
    \centering
	\includegraphics[width=\bigImageSize]{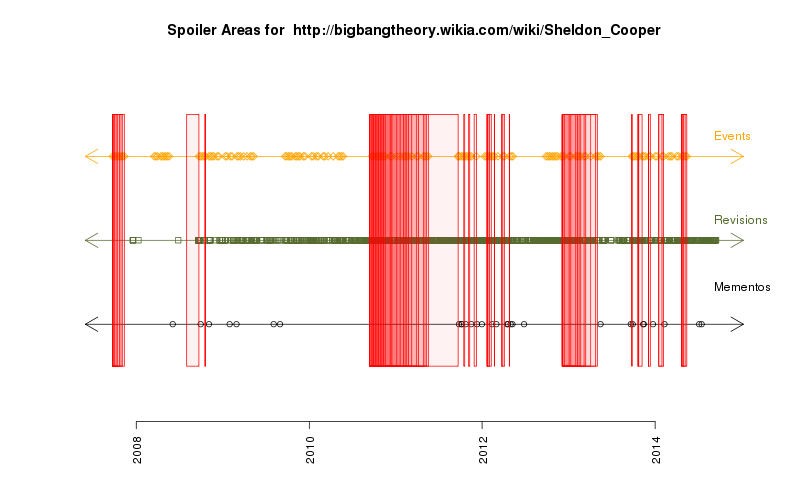}
	\caption{Spoiler areas for the page in \emph{the Big Bang Theory Wiki} that contains the most revisions\protect\footnotemark}
	\label{fig:spoiler-areas-Sheldon-Cooper}
\end{figure*}
\footnotetext{\url{http://bigbangtheory.wikia.com/wiki/Sheldon_Cooper}}

\clearpage
\begin{figure*}[!htp]
    \centering
	\includegraphics[width=\bigImageSize]{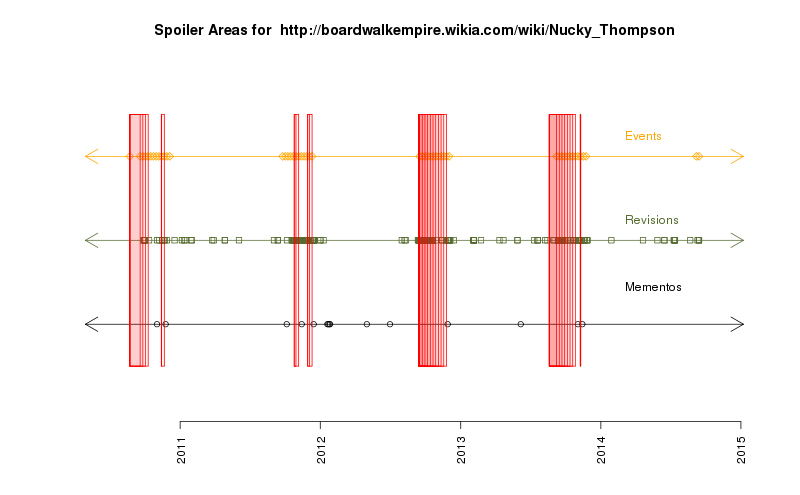}
	\caption{Spoiler areas for the page in the \emph{Boardwalk Emprire Wiki} that contains the most revisions\protect\footnotemark}
	\label{fig:spoiler-areas-Nucky-Thompson}
\end{figure*}
\footnotetext{\url{http://boardwalkempire.wikia.com/wiki/Nucky_Thompson}}

\begin{figure*}[!htp]
    \centering
	\includegraphics[width=\bigImageSize]{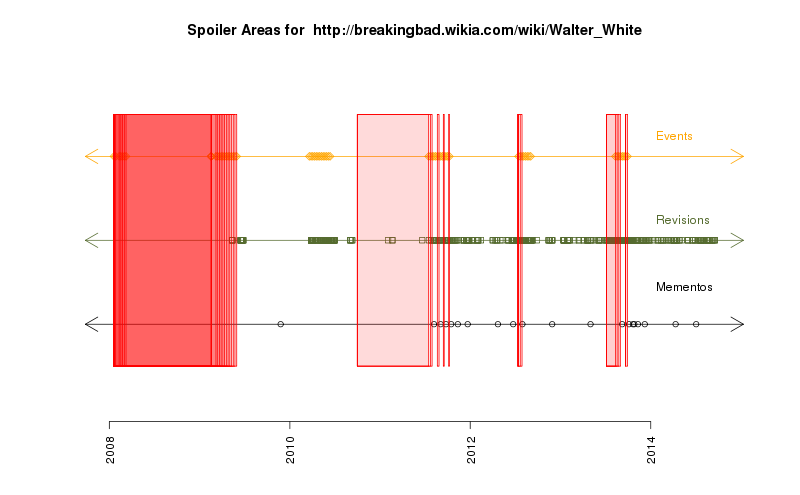}
	\caption{Spoiler areas for the page in the \emph{Breaking Bad Wiki} that contains the most revisions\protect\footnotemark}
	\label{fig:spoiler-areas-Walter-White}
\end{figure*}
\footnotetext{\url{http://breakingbad.wikia.com/wiki/Walter_White}}

\clearpage
\begin{figure*}[!htp]
    \centering
	\includegraphics[width=\bigImageSize]{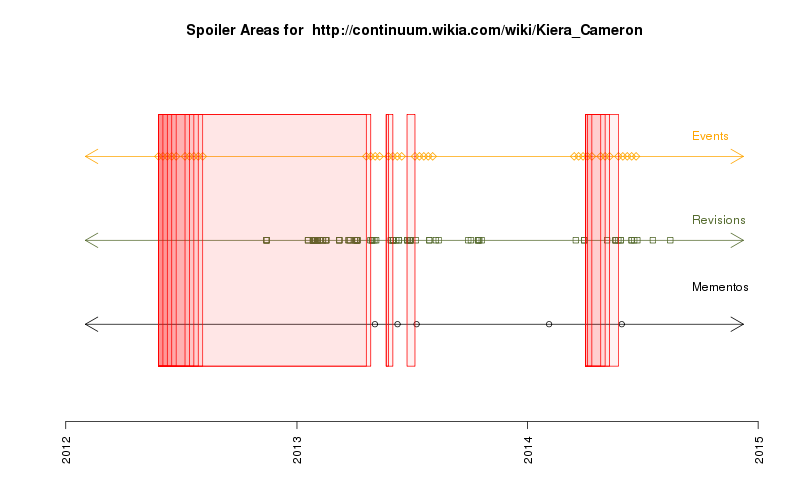}
	\caption{Spoiler areas for the page in the \emph{Continuum Wiki} that contains the most revisions\protect\footnotemark}
	\label{fig:spoiler-areas-Kiera-Cameron}
\end{figure*}
\footnotetext{\url{http://continuum.wikia.com/wiki/Kiera_Cameron}}

\begin{figure*}[!htp]
    \centering
	\includegraphics[width=\bigImageSize]{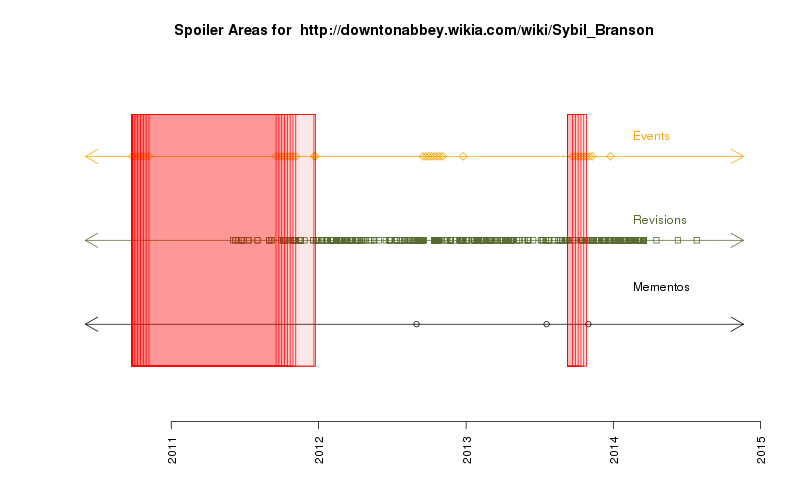}
	\caption{Spoiler areas for the page in the \emph{Downton Abbey Wiki} that contains the most revisions\protect\footnotemark}
	\label{fig:spoiler-areas-Sybil-Branson}
\end{figure*}
\footnotetext{\url{http://downtonabbey.wikia.com/wiki/Sybil_Branson}}

\clearpage
\begin{figure*}[!htp]
    \centering
	\includegraphics[width=\bigImageSize]{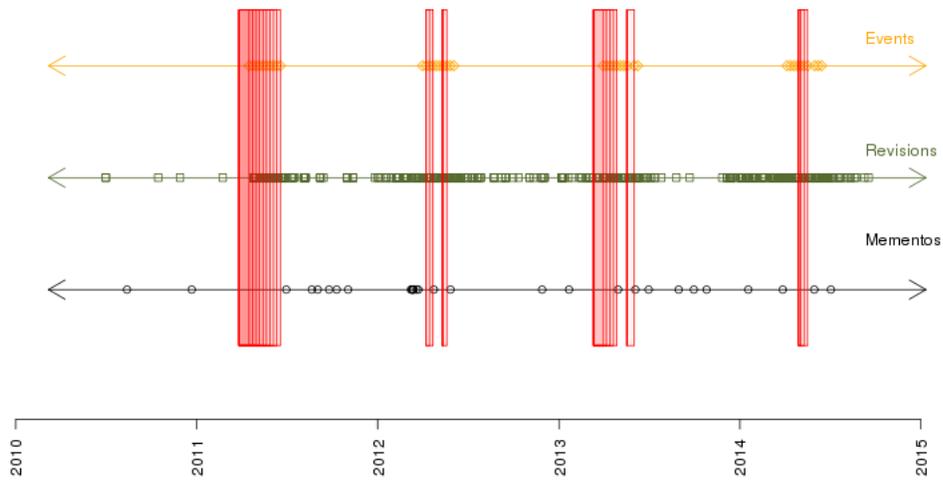}
	\caption{Spoiler areas for the most popular page in the \emph{Game of Thrones Wiki}\protect\footnotemark}
	\label{fig:spoiler-areas-Daenerys-Targaryen-2}
\end{figure*}
\footnotetext{\url{http://gameofthrones.wikia.com/wiki/Daenerys_Targaryen}}

\begin{figure*}[!htp]
    \centering
	\includegraphics[width=\bigImageSize]{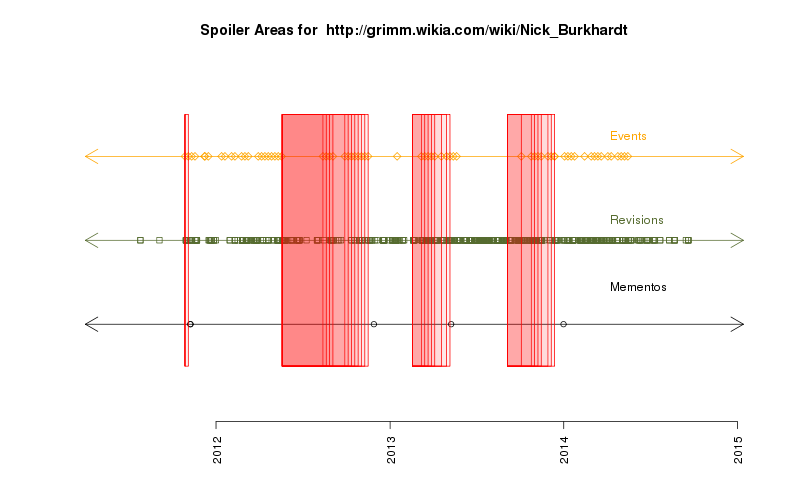}
	\caption{Spoiler areas for the page in the \emph{Grimm Wiki} that contains the most revisions\protect\footnotemark}
	\label{fig:spoiler-areas-Nick-Burkhardt}
\end{figure*}
\footnotetext{\url{http://grimm.wikia.com/wiki/Nick_Burkhardt}}

\clearpage
\begin{figure*}[!htp]
    \centering
	\includegraphics[width=\bigImageSize]{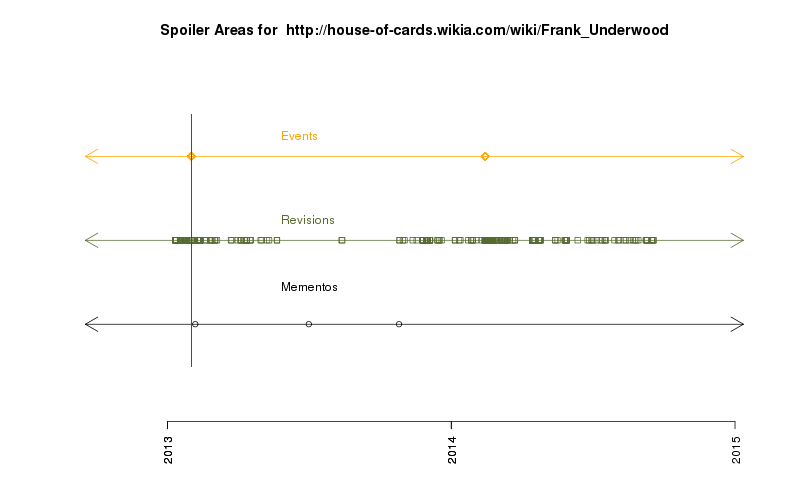}
	\caption{Spoiler areas for the most popular page in the \emph{House of Cards Wiki}\protect\footnotemark}
	\label{fig:spoiler-areas-Frank-Underwood-2}
\end{figure*}
\footnotetext{\url{http://house-of-cards.wikia.com/wiki/Frank_Underwood}}

\begin{figure*}[!htp]
    \centering
	\includegraphics[width=\bigImageSize]{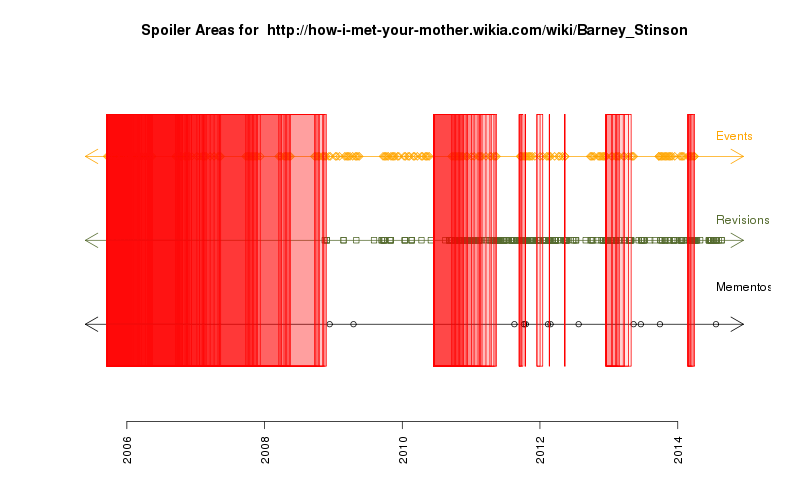}
	\caption{Spoiler areas for the most popular page in the \emph{How I Met Your Mother Wiki}\protect\footnotemark}
	\label{fig:spoiler-areas-Barney-Stinson}
\end{figure*}
\footnotetext{\url{http://how-i-met-your-mother.wikia.com/wiki/Barney_Stinson}}

\clearpage
\begin{figure*}[!htp]
    \centering
	\includegraphics[width=\bigImageSize]{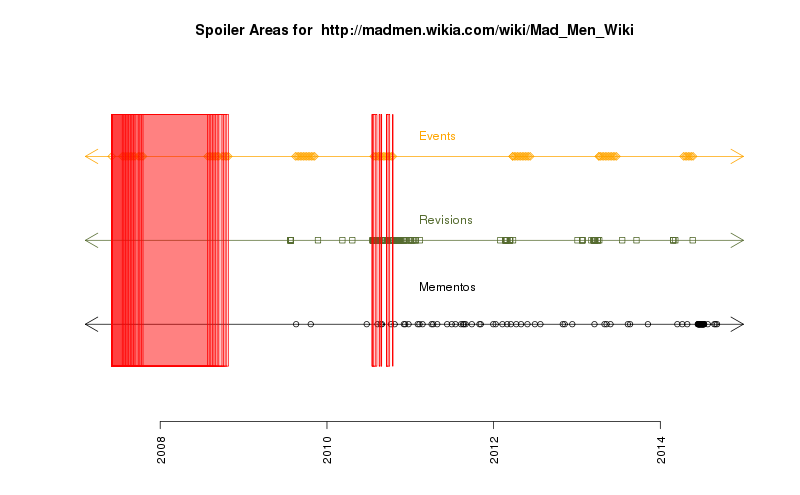}
	\caption{Spoiler areas for the page in the \emph{Mad Men Wiki} that contains the most revisions\protect\footnotemark}
	\label{fig:spoiler-areas-Mad-Men-Wiki}
\end{figure*}
\footnotetext{\url{http://madmen.wikia.com/wiki/Mad_Men_Wiki}}

\begin{figure*}[!htp]
    \centering
	\includegraphics[width=\bigImageSize]{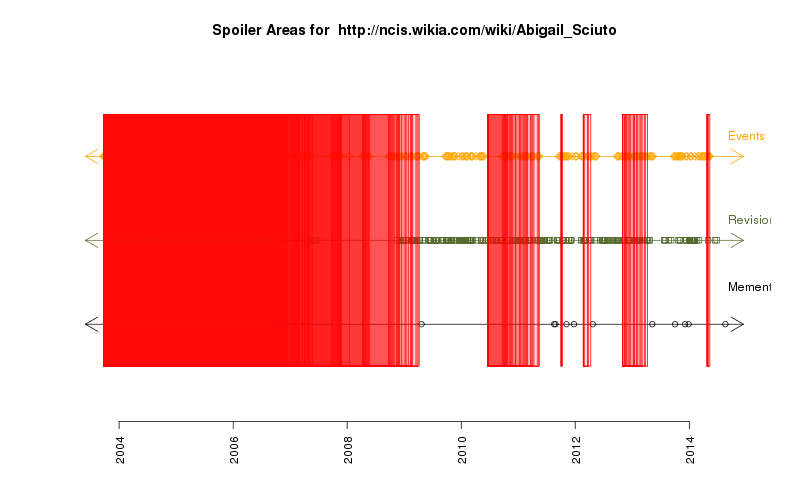}
	\caption{Spoiler areas for the page in the \emph{NCIS Database} that contains the most revisions\protect\footnotemark}
	\label{fig:spoiler-areas-Abigail-Sciuto}
\end{figure*}
\footnotetext{\url{http://ncis.wikia.com/wiki/Abigail_Sciuto}}

\clearpage
\begin{figure*}[!htp]
    \centering
	\includegraphics[width=\bigImageSize]{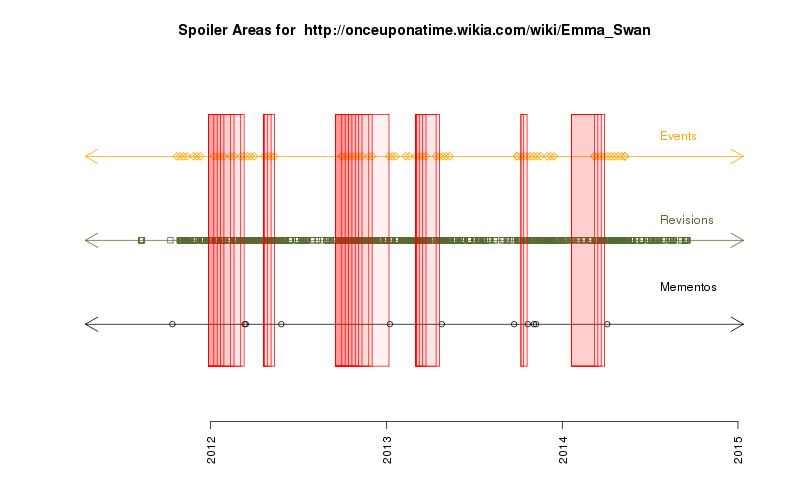}
	\caption{Spoiler areas for the page in the \emph{Once Upon A Time Wiki} that contains the most revisions\protect\footnotemark}
	\label{fig:spoiler-areas-Emma-Swan}
\end{figure*}
\footnotetext{\url{http://onceuponatime.wikia.com/wiki/Emma_Swan/Gallery}}

\begin{figure*}[!htp]
    \centering
	\includegraphics[width=\bigImageSize]{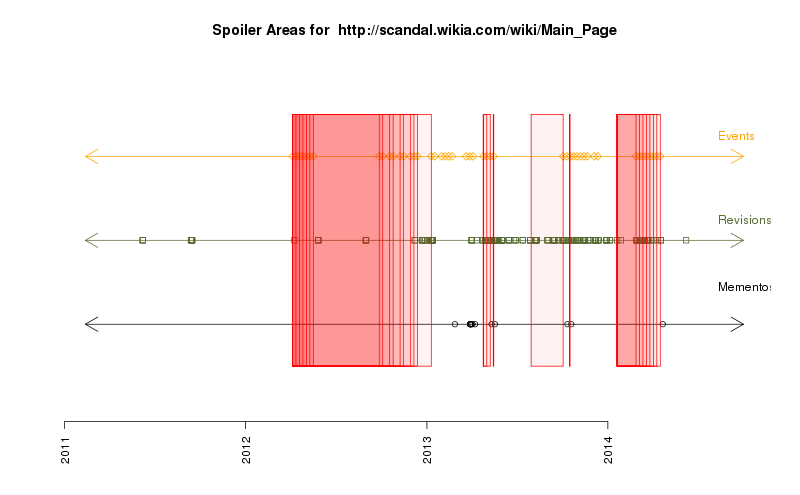}
	\caption{Spoiler areas for the page in the \emph{Scandal Wiki} that contains the most revisions\protect\footnotemark}
	\label{fig:spoiler-areas-Scandal-Main-Page}
\end{figure*}
\footnotetext{\url{http://scandal.wikia.com/wiki/Main_Page}}

\clearpage
\begin{figure*}[!htp]
    \centering
	\includegraphics[width=\bigImageSize]{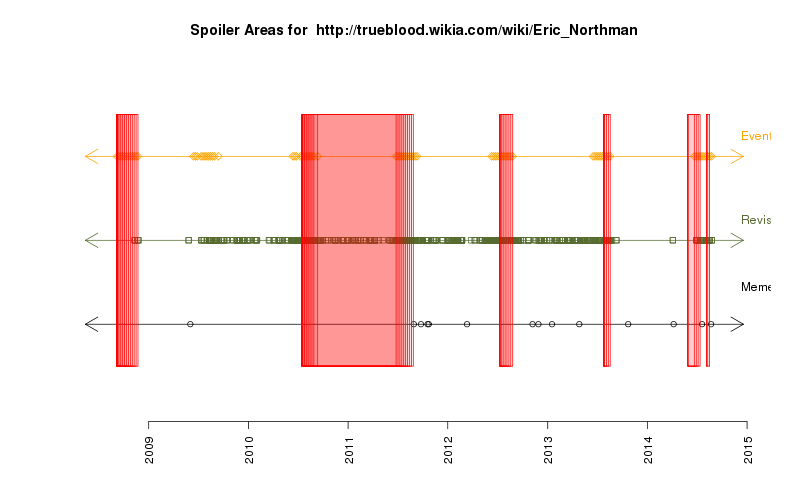}
	\caption{Spoiler areas for the page in the \emph{True Blood Wiki} that contains the most revisions\protect\footnotemark}
	\label{fig:spoiler-areas-Eric-Northman}
\end{figure*}
\footnotetext{\url{http://trueblood.wikia.com/wiki/Eric_Northman}}

\begin{figure*}[!htp]
    \centering
	\includegraphics[width=\bigImageSize]{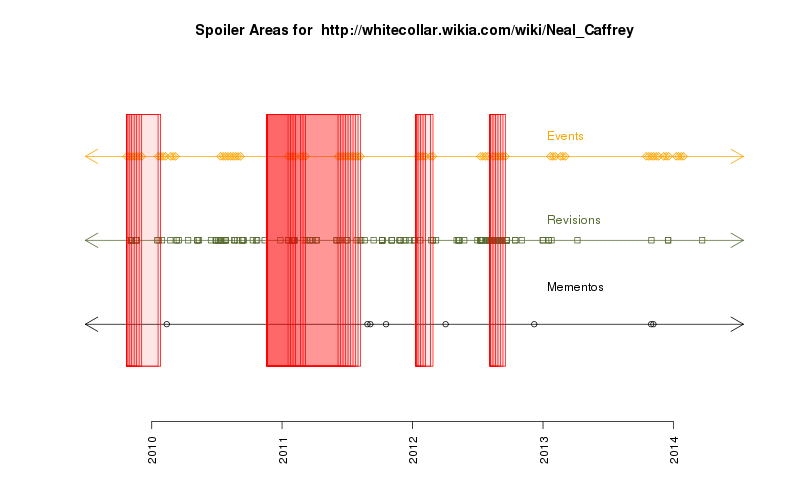}
	\caption{Spoiler areas for the page in the \emph{White Collar Wiki} that contains the most revisions\protect\footnotemark}
	\label{fig:spoiler-areas-Neal-Caffrey}
\end{figure*}
\footnotetext{\url{http://whitecollar.wikia.com/wiki/Neal_Caffrey}}

%\balancecolumns

\end{document}